\pdfoutput=1
\documentclass{webofc}
\usepackage[varg]{txfonts}   
\usepackage[colorlinks,hyperindex, 
           bookmarks=true,bookmarksopen=true]{hyperref} 
\hypersetup{                 
    colorlinks=true,         
    linkcolor=blue,          
    citecolor=red,           
    urlcolor=cyan          
}

\begin{document}
\title{Decays of the tensor glueball in a chiral approach}
%
%

\author{\firstname{Arthur} \lastname{Vereijken}\inst{1}\fnsep\footnote{\href{mailto:arthur.vereijken@gmail.com}{arthur.vereijken@gmail.com}}
}

\institute{Institute of Physics, Jan Kochanowski University, ul. Uniwersytecka 7, 25-406, Kielce, Poland}

\abstract{Glueballs remain an experimentally undiscovered prediction of QCD. Lattice QCD predicts a spectrum of glueballs, with the tensor $(J^{PC}=2^{++})$ glueball being the second lightest, behind the scalar glueball. From an effective hadronic model based on spontaneous and explicit chiral symmetry breaking, we compute decay ratios of the tensor glueball into various meson decay channels. We find the tensor glueball to primarily decay into 2 vector mesons, dominated by $\rho\rho$ and $K^*K^*$. These results are compared to experimental data of decay rates of spin 2 mesons. Based on this comparison we make statements on the eligibility of these mesons as potential tensor glueball candidates.}
\maketitle
\section{Introduction}
\label{intro}
The experimental verification of glueballs has been, and still is, a long-standing open issue in QCD \cite{Gross:2022hyw}. Numerous theoretical \cite{Giacosa:2005bw,Brunner:2015oqa} and experimental \cite{Klempt:2022qjf} approaches have made headway, yet the situation is still not completely clear \cite{Mathieu:2008me,Crede:2008vw,Llanes-Estrada:2021evz,Chen:2022asf}. The different theoretical methods agree on the mass hierarchy of the lowest lying glueball states, with the scalar $(J^{PC}=0^{++})$ being the lightest and the tensor $(J^{PC}=2^{++})$ the second lightest glueball. In this work we will focus on the tensor glueball, for which there are many experimentally observed isoscalar-tensor candidate resonances. We will present results on the tensor glueball \cite{Vereijken:2023jor} in the extended Linear Sigma Model \cite{Parganlija:2012fy}, which is an extension of earlier works on axial-tensor mesons in the same model \cite{Jafarzade:2022uqo}. Different glueballs have been studied before in the same type of model, such as the scalar \cite{Janowski:2014ppa} and the pseudoscalar glueball\cite{Eshraim:2012jv}.

\section{Chiral model}
\label{sec-1}
The meson resonances are gathered into the nonets $V^{\mu}$ $(J^{PC}=1^{--})$ containing vector mesons, $A_{1}^{\mu}$ $(J^{PC}=1^{++})$ containing axial-vector mesons, $P$ $(J^{PC}=0^{-+})$ containing pseudoscalar mesons, $S$ $(J^{PC}=0^{++})$ containing scalar mesons, $T^{\mu\nu}$ $(J^{PC}=2^{++})$ containing tensor mesons, and $A_{2}^{\mu\nu}$ $(J^{PC}=2^{--})$ containing axial-tensor mesons. For details on the resonance assignment of the nonets see \cite{Vereijken:2023jor,Jafarzade:2022uqo}. The tensor glueball itself is a flavor blind object $G_{2,\mu\nu}$.
\\
\\
The chiral invariant Lagrangians relevant to us are as follows
\begin{align}\label{lag1}
    \mathcal{L}_{\lambda}=\frac{\lambda}{\sqrt{6}} G_{2,\mu\nu}\Big(\text{Tr}\Big[ \{ L^{\mu}, L^{\nu}\}\Big]+\text{Tr}\Big[ \{R^{\mu}, R^{\nu}\} \Big]\Big) 
    \text{ ,}
\end{align}
\begin{align}\label{lag2}
    \mathcal{L}_{\alpha}=\frac{\alpha}{\sqrt{6}} G_{2,\mu\nu}\Big(\text{Tr}\Big[ \Phi \textbf{R}^{\mu\nu}\Phi^\dagger\Big]+\text{Tr}\Big[ \Phi^\dagger\textbf{L}^{\mu\nu}\Phi\Big]\Big)  
    \text{ .}
\end{align}
These are the leading terms in large-$N_c$ expansion, where $N_c$ is the number of colors of underlying the gauge group. The nonets of chiral partners are grouped together and are given by:
\begin{align}
     L^{\mu}&:= V^{\mu}+A_1^{\mu} \ \text{ , }  R^{\mu}:= V^{\mu}-A_1^{\mu} \text{ ,  } \Phi = S + i P \nonumber 
     \\
      \mathbf{L}^{\mu\nu}&=T^{\mu \nu}+A_2^{\mu \nu} \text{ , 
 }\mathbf{R}^{\mu\nu}=T^{\mu \nu}-A_2^{\mu \nu},
\end{align}
such that they obey the transformation rules $L^{\mu} \to U_L L^{\mu}U_L^{\dagger} $, $ R^{\mu} \to U_{R}R^{\mu}U^{\dagger}_{R}$, $\Phi \to U_L\Phi U^{\dagger}_R , \mathbf{R}^{\mu\nu} \to U_{R}\mathbf{R}^{\mu\nu}U^{\dagger}_{R} ,\mathbf{L}^{\mu\nu} \to U_{L}\mathbf{L}^{\mu\nu}U^{\dagger}_{L}$ under the chiral transformations of $U_L(3) \times U_R(3)$. The first Lagrangian \eqref{lag1} models the 2 body decays of the tensor glueball into 2 vector mesons, into 2 pseudoscalar mesons, and into an axial-vector and a pseudoscalar meson. The second Lagrangian \eqref{lag2} leads to the decay into a tensor and a pseudoscalar meson. Since the coupling constants $\alpha$ and $\lambda$ are not a priori known and cannot be fitted to experimental data, we are limited to computing decay ratios, seperately for each Lagrangian. Lattice calculations in \cite{Athenodorou:2020ani} find a tensor glueball mass of $2369$ MeV. The decay ratios of the first Lagrangian with respect to $\pi\pi$ for this mass are shown in table \ref{tab:results}. As evident from the results in table \ref{tab:results}, the 2-vector decay channel is dominant, in particular the decays into $\rho\rho$ and $K^{*}\bar{K}^{*}$. A similar dominance of the 2-vector channel was recently found in \cite{Hechenberger:2023ljn} with the holographic Witten-Sakai-Sugimoto model.

\begin{table}[ptb]
\centering
\renewcommand{\arraystretch}{2.} 
\begin{tabular}{|c|c|c|c|c|c|}
\hline
Decay Ratio   & theory & Decay Ratio   & theory &Decay Ratio   & theory \\ \hline\hline
	$ \frac{G_{2}(2369) \longrightarrow \bar{K}\, K}{G_{2}(2369) \longrightarrow  \pi\,\pi} $ & $0.4$ & $ \frac{G_{2}(2369) \longrightarrow \rho(770)\, \rho(770)}{G_{2}(2369) \longrightarrow  \pi\,\pi} $ & $51$ & $ \frac{G_{2}(2369)\longrightarrow a_1(1260)\,\pi}{G_{2}(2369) \longrightarrow  \pi\, \pi} $ & $0.26$ 
\\ \hline
 $ \frac{G_{2}(2369) \longrightarrow \eta \, \eta}{G_{2}(2369) \longrightarrow  \pi\,\pi}$  & $0.1$ & $ \frac{G_{2}(2369) \longrightarrow \bar{K}^\ast(892)\, \bar{K}^\ast(892)}{G_{2}(2369) \longrightarrow  \pi\,\pi} $ & $44$ & $ \frac{G_{2}(2369)\longrightarrow K_{1\,,A}\,K}{G_{2}(2369)\longrightarrow \pi\,\pi} $ & $0.12$
\\ \hline
$ \frac{G_{2}(2369) \longrightarrow \eta \, \eta^\prime}{G_{2}(2369) \longrightarrow  \pi\,\pi}$  & $0.005$ & $ \frac{G_{2}(2369) \longrightarrow \omega(782) \, \omega(782)}{G_{2}(2369) \longrightarrow  \pi\,\pi}$  & $17$ & $ \frac{G_{2}(2369)\longrightarrow f_1(1285)\,\eta}{G_{2}(2369)\longrightarrow \pi\,\pi}$  & $0.03$
\\ \hline
$ \frac{G_{2}(2369) \longrightarrow \eta^\prime \, \eta^\prime}{G_{2}(2369) \longrightarrow  \pi\,\pi}$  & $0.01$ &
$ \frac{G_{2}(2369) \longrightarrow \phi(1020) \,\phi(1020)}{G_{2}(2369) \longrightarrow  \pi\,\pi}$  & $7$ &	$ \frac{G_{2}(2369)\longrightarrow f_1(1420)\,\eta}{G_{2}(2369)\longrightarrow \pi\,\pi}$  & $0.008$
\\ \hline
\end{tabular}%
\caption{Decay ratios of  $G_2$ w.r.t. $\pi \pi$ for a mass of $2369$ MeV. The columns are sorted as $PP$ on the left, $VV$ in the middle, and $AP$ on the right.}\label{tab:results}
\end{table}

\section{Results \& Data Comparison}

\begin{table}[ht]
\centering
\begin{tabular}{|c|c|c|c|}
\hline
Resonances & Decay Ratios       & PDG    \cite{Workman:2022ynf}             & Model Prediction 
\\ \hline \hline
$f_2(1910)$   & $\rho(770)\rho(770) / \omega(782)\omega(782)$ &   $2.6\pm 0.4$ &    $3.1$        \\ 
\cline{1-4}
 \hline
$f_2(1910)$   & $f_2(1270)\eta/a_2(1320)\pi$ &   $0.09\pm 0.05$ &    $0.07$       \\ 

\cline{1-4}
 \hline
$f_2(1910)$   & $\eta \eta / \eta\eta^\prime(958)$ &   $<0.05$ &    $\sim 8$       \\ 

\cline{1-4}
 \hline
$f_2(1910)$   & $\omega(782) \omega(782) /\eta \eta\prime(958)$ &   $2.6 \pm 0.6$  &    $\sim 200$       \\ 
\hline \hline
\cline{1-4}
$f_2(1950)$   & $\eta\eta/\pi\pi$ &     $0.14 \pm 0.05$                  & $0.081$                       \\ 
\cline{1-4}
\hline
$f_2(1950)$   & $K\overline{K}/\pi\pi$ &     $\sim 0.8$                  & $0.32$                       \\ 
\cline{1-4}
\hline
$f_2(1950)$   & $4\pi/\eta\eta$ &     $>200$                  & $>700$                       \\ 

\hline \hline
\cline{1-4}
\hline
$f_2(2150)$   & $f_2(1270)\eta/a_2(1320)\pi$ &     $0.79\pm 0.11$                  & $0.1$                       \\ 
\cline{1-4}
\hline
$f_2(2150)$   & $K\overline{K} / \eta \eta$ &     $1.28\pm 0.23$                 & $\sim 4$                       \\ 
\cline{1-4}
\hline
$f_2(2150)$   & $\pi \pi / \eta \eta$ &     $<0.33$                 & $\sim 10$                       \\ 
\cline{1-4}
\hline \hline
$f_J(2220)$   & $\pi \pi / K\overline{K}$ &     $1.0 \pm 0.5$                 & $\sim 2.5$                       \\ 
\hline
\end{tabular}
\caption{Decay ratios for the decay channels with available data.}
\label{BR-table}
\end{table}
We compare results to available data of spin-2 isoscalar resonances ($J^{PC}=2^{++}$, I = 0) with masses of 1.9 GeV and upwards. These are the $f_{2}(1910), f_{2}(1950), f_{2}(2010), f_{2}(2150), f_{J}(2220), f_{2}(2300),$ and the $f_{2}(2340)$. In table \ref{BR-table} decay ratios are computed and compared with PDG data \cite{Workman:2022ynf} where available, revealing how well they fit as glueball candidates. We see that every glueball candidate other than the $f_{2}(1950)$ has disagreement with experimental data, and that the $f_{2}(1950)$ fits reasonably well, given uncertainties on both sides. Therefore we interpret $f_{2}(1950)$ to be the best candidate for the lightest tensor glueball, which has not been the first time it has been proposed as the tensor glueball, see e.g. \cite{Godizov:2016vuw}.

\section{Conclusion}
In this note we have computed decays of the tensor glueball using a chiral hadronic model. We find that the decay to two vectors, in particular $\rho \rho$ and $K^{*}\bar{K}^{*}$, is dominant. Upon comparing with experimental data, we find that the $f_{2}(1950)$ is the most suitable candidate (even if some deviations are present) for the tensor glueball based on the known decay ratios. In the future, one should investigate why there is up to a $400$ MeV mass difference between this resonance and predicted masses from lattice methods. One possible explanation could be the role of mesonic loops and/or the mixing with nearby quark-antiquark states.

\section*{Acknowledgements}
We thank Francesco Giacosa and Shahriyar Jafarzade for useful discussions. We also acknowledge financial support from the Polish National Science Centre (NCN) via the OPUS project 2019/33/B/ST2/00613.

%
%
%

\end{document}